# Research on dynamic analysis and prediction model of tennis match based on Bayesian probability and analytic Hierarchy process

Chuangqi Li*[a]

[a]College of Computer and Information Engineering, Henan Normal University

In the 2023 Wimbledon Gentlemen's final, Carlos Alcaraz defeated Novak Djokovic. This study aims to predict athletes' performance through five key aspects: first, a multi-classification model based on logistic regression was established, yielding probabilities of winning and losing the first serve at 0.6734 and 0.3266, respectively, and validated with match 1701. Second, an unsupervised "momentum" evaluation model using AHP analytic hierarchy process showed a strong correlation between "momentum" score and winning rate. Third, a trend analysis model identified psychological factors as significantly impacting results. Fourth, the model's generalization was tested with additional competition data, including women's tennis matches. Finally, data analysis suggested that coaches should focus on improving mental toughness and serve-receive skills, as these significantly affect momentum.

# Contents

# 1 Introduction

## 1.1 Background

In the 2023 Wimbledon Gentlemen's final, 20-year-old Spanish rising star Carlos Alcaraz defeated 36-year-old Novak Djokovic. The loss was Djokovic's first at Wimbledon since 2013 and ended a remarkable run for one of the all-time great players in **Grand Slams**.

We can see all the points for the first set when Djokovic had the edge using the "set_no" column equal to 1. The incredible swings, sometimes for many points or even games, that occurred in the player who seemed to have the advantage are often attributed to "momentum."

## 1.2 Restatement of the Problem

• Develop a model that captures the flow of play as points occur and apply it to one or more of the matches.
• A tennis coach is skeptical that "momentum" plays any role in the match. Instead, he postulates that swings in play and runs of success by one player are random. Use your model/metric to assess this claim.
• Using the data provided for at least one match, develop a model that predicts these swings in the match. What factors seem most related (if any)?
• Given the differential in past match "momentum" swings how do you advise a player going into a new match against a different player?
• Test the model you developed on one or more of the other matches.

## 1.3 Our Work

We were asked to develop a model to capture the flow of a tennis game when a point is scored, study the role of "momentum" in the game, check the reliability of the model, and test the generality of the model to other games. Based on this, we worked as follows:

• Data preprocessing, cleaning and reorganizing the data. A multi-classification model based on logistic regression is established, and a multi-time scale winning rate classification prediction model is constructed by Bayesian probability. Model accuracy was evaluated by calculating macro and micro F1 scores. The 1701 match is used to verify the validity of the model, which reflects the superiority of the model.

• Delved into how 'momentum' is calculated, breaking it down into the sum of scores over different time periods. An unsupervised "momentum" evaluation model based on AHP analytic hierarchy process with subjective evaluation is constructed. The strong correlation between "momentum" and win rate is confirmed by 3D surface trend fitting and cosine similarity calculation.

• Through the trend analysis model, it is observed that when the strength of the two sides in the game is similar, the momentum shows a trend of reversal of fluctuations. Based on the network search method, the momentum fluctuation prediction model of the player under the condition of successively serving is established. The momentum amplitude time-frequency map is obtained by wavelet transform.

• The generalization ability of the model was evaluated in depth, and the data of other competitions were introduced and optimized. The data related to women's tennis match was introduced to realize the mobility prediction of the model.
• The data analysis highlights the significant impact of the serving session on momentum. Recommendations were made that coaches should focus on the improvement of mental toughness, serving and receiving skills to adapt to the changes in momentum during the match.

## 2 Assumptions and Justifications

To simplify our model and eliminate the complexity, we make the following main assumptions in this paper.

**Assumption.1:** The service results of each player are not affected by the conditions of the court, their own equipment and other aspects, and the strength of each player is almost the same.

**Assumption.2:** The result of the serve is only affected by small changes in the player's own state, such as fatigue level, excitement level, technical state, etc.

**Assumption.3:** These small changes will gradually accumulate in the course of the game, and eventually cause the game result to change. The result of the game can be predicted by monitoring these small changes[1][2].

## 3 Notations

The primary notations used in this paper are listed in Table 1.

**Table 1: Notations**

| Symbol | Description |
|---|---|
| P(A|B) | The probability of A conditional on B |
| TPR | True Positive Rate |
| FPR | False Positive Rate |
| $F1_{micro}$ | The value calculated using F1 calculation formula |
| $P_{macro}$ | Average precision |
| $R_{macro}$ | Average recall |
| $w_i$ | The weight of the competition |
| $P(s)$ | Momentum score in the first round |
| $M(s)$ | The score of the first and last three rounds of the game |
| $N(s)$ | Scores in the previous seven games |
| $W(a,b)$ | momentum score signal |
| $f(t)$ | Wavelet transform coefficients in scale (frequency dependent) and displacement (time dependent) |
| $a$ | the scale parameter |
| $b$ | the displacement parameter |
| $\psi^*(t)$ | represents the complex conjugate of the wavelet basis function |

# 4 Problem 1: Multi-time scale winning rate classification prediction model based on Bayesian probability

We built multiple models based on different time scales and feature dimensions, using round, round and plate as the scale statistical data respectively. Feature engineering was performed and important features were extracted as input variables. Using existing data, the Bayesian probability of a player winning in the case of a serve is calculated and used as the output. A Bayesian-based logistic regression multi-class prediction model was used to compare the performance of models with different time scales and whether or not Bayesian probability was introduced on generalization indicators such as classification accuracy. Finally, we verified these models on 1701 specific game data to obtain the changes in winning and losing in the game and the corresponding probability values[3].

## 4.1 Data Preprocessing

Based on a brief browse of the formal text and image data, we identified three issues:
• Consider bad pixel data
• Abnormal data exists
• Missing data exists

In order to solve the problems existing in these data and build a valid data set, we took the following measures:

**Step 1:** Set the negative value of AD to any number greater than 40, here we take 50. It may be necessary to determine the outcome of each game by judging the size of p1_score and p2_score, or to judge the intensity of the game through Boolean values.

**Step 2:** Perform average filling and replacement processing of missing values, bad pixel data and other data that violate the rules of tennis matches.

**Step 3:** Convert categorical variables into continuous quantitative variables (for example: F (Forebad) becomes 1, B (Backhand) becomes 2).

**Step 4:** On this basis, the internal connections of the data are mined, and statistics are collected on the number of points or games won in a row, the change in the ratio of unforced errors to winning points in the game, and signs of the players' physical condition, and used as new feature data for model training.

**Step 5:** Summarize the data of Player 1 and Player 2, observe the overall change trend, and integrate the data according to different time scales to generate a heat map of the correlation between the rounds, points, games and sets won by Player 1 and Player 2. Analysis, as shown in Figure 1.

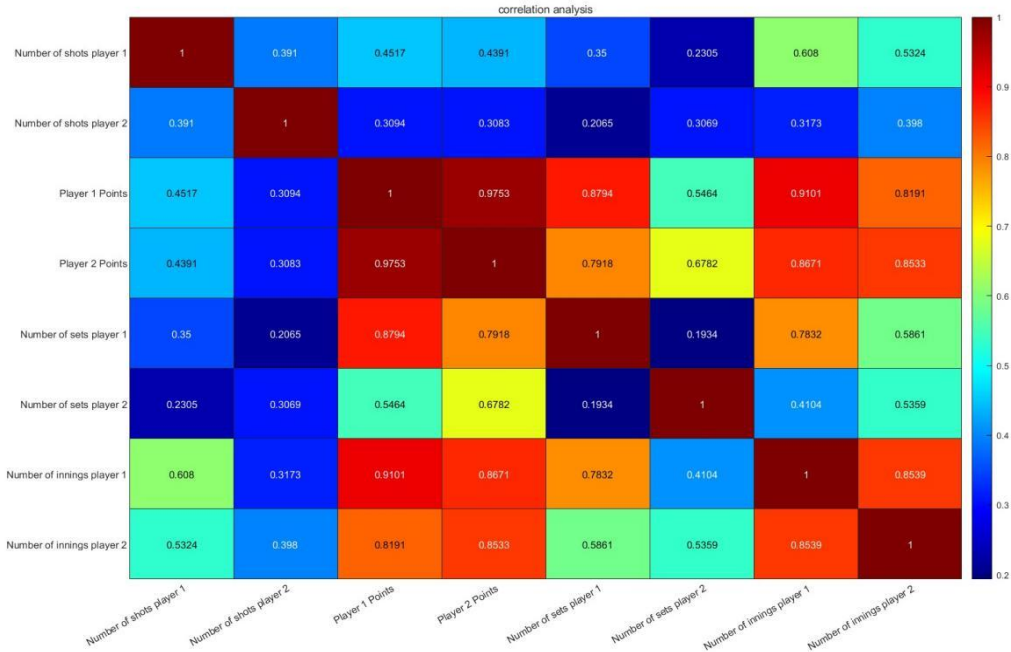

**Figure 1** Correlation analysis of wins between player 1 and player 2

All in all, there is a certain correlation between Player 1 and Player 2 at different time scales, which provides data support for subsequent model construction and cross-validation.

## 4.2 The Establishment of The Model

Considering Bayesian probability, calculate the probability of the first server winning in each round. The specific steps are as follows:

**Step 1:** Bayesian probability calculation: Based on historical data, set the initial winning probability of each player. The Bayesian formula is used to update the player's winning probability by combining information such as the first server and the game score.

**Step 2:** Feature Engineering: Extract features related to game dynamics from the original data, including winning probability after Bayesian update.

**Step 3:** Construct a multi-time scale logistic regression multi-classification model: integrate game data sets at different time scales (rounds/rounds/sets), and estimate model parameters through the maximum likelihood method.

**Step 4:** Network training and model parameter adjustment: Use the training set for network training, and adjust the weights through the backpropagation algorithm to minimize the prediction error. Adjust the model structure or training strategy based on the evaluation results to improve prediction accuracy.

**Step 5:** Describe the players' game performance at different time scales: calculate the comparative changes of each player's probability of winning the first serve and the probability of winning the game under different models, and display them through visualization.

### 4.2.1 Bayesian probability

A model that incorporates the advantage of serving first into the Bayesian probability model: First, assume that $P(W)$ is the prior probability of any player winning the game when player 1 serves first. If the player serves first, the $P(W)$ of this information $B$ is $P(W|B)$, which is the posterior probability of winning the game if it is known that the player served first.

$$P(W|B) = \frac{P(B|W) \cdot P(W)}{P(B)} \quad (1)$$

$P(W|B)$ is the posterior probability of winning the game under the condition that the player serves first.

$P(B|W)$ is the probability of being the first server under the condition that the player wins the game.

$P(W)$ is the probability that the player wins the game The prior probability can be estimated based on historical data.

$P(B)$ is the probability of any player serving first, which is a constant.

On this basis, by adjusting $P(W)$ to reflect the advantage of serving first, a more accurate prediction can be made. Match Results. We calculated the probability of winning if player 1 serves first and player 2 serves first:

$$P(\text{Player 1 wins}|\text{Player 1 starts}) = \frac{P(\text{Player 1 wins} + \text{Player 1 starts})}{P(\text{Player 1 starts})} \quad (2)$$

$$P(\text{Player 2 wins}|\text{Player 2 starts}) = \frac{P(\text{Player 2 wins} + \text{Player 2 starts})}{P(\text{Player 2 starts})} \quad (3)$$

The probability of failure is $1 - P$. Finally, the average probability of player 1 and player 2 winning is 0.6734, and the probability of failure is 0.3266. We replace this probability with the output variable $\{0,1\}$ The sample output label corresponding to the first serve in the set.

### 4.2.2 Logistic regression multi-classification model

The expression of the logical function (Sigmoid function) is as follows:

$$S(z) = \frac{1}{1 + e^{-z}} \quad (4)$$

where z is the linear combination $\beta_0 + \beta_1 X_1 + \beta_2 X_2 + \cdots + \beta_P X_P$.

The training goal of the logistic regression model is to find the best parameters $\beta_1, \beta_2, \beta_3, \cdots, \beta_P$ to make the model produce the best classification prediction. Typically, the training process uses methods such as maximum likelihood estimation to estimate these parameters.

Logistic regression can be expanded from binary classification problems to multi-classification problems.

$$\log it(A) = \log \frac{P(Y = A)}{P(Y = C)} = a_0 + a_1 x \quad (5)$$

$$\log it(B) = \log \frac{P(Y = B)}{P(Y = C)} = b_0 + b_1 x \quad (6)$$

Therefore

$$P(Y = A) = \frac{e^{a_0 + a_1 x}}{1 + e^{a_0 + a_1 x} + e^{b_0 + b_1 x}} \quad (7)$$

$$P(Y = B) = \frac{e^{a_0 + b_1 x}}{1 + e^{a_0 + a_1 x} + e^{b_0 + b_1 x}} \quad (8)$$

$$P(Y = C) = 1 - P(Y = A) - P(Y = B) = \frac{1}{1 + e^{a_0 + a_1 x} + e^{b_0 + b_1 x}} \quad (9)$$

## 4.3 Model accuracy test and prediction

In the learning of traditional classification models, the generalization ability of the model is evaluated through classification accuracy, recall, and precision. At the same time, considering that the classification result after calculating the winning rate through Bayesian is greater than Category 2 (category 4) is a multi-classification problem, so macro F1 score and micro F1 score are introduced to calculate a comprehensive score. The formula is as follows:

Classification Accuracy, Recall (R), Precision (P)

$$Accuracy = \frac{TP + TN}{TP + TN + FP + FN}, \quad R = \frac{TP}{TP + FN}, \quad P = \frac{TP}{TP + FP} \quad (10)$$

Macro F1 score and micro F1 score MicroF1 score, when each category is regarded as a positive category, a set of TP, FP and FN values can be calculated. We sum them up to calculate TP, FP and FN for all samples, and then Applying the formula of F1 score again, P and R of category i can be expressed as:

$$P_{micro} = \frac{\sum_{i=1}^{n} TP_i}{\sum_{i=1}^{n} TP_i + \sum_{i=1}^{n} FP_i} \quad R_{micro} = \frac{\sum_{i=1}^{n} TP_i}{\sum_{i=1}^{n} TP_i + \sum_{i=1}^{n} FN_i} \quad (11)$$

Then use the F1 calculation formula to calculate the F1 value, which is Micro-F1:

$$F1_{macro} = 2 \cdot \frac{P_{macro} \cdot R_{macro}}{P_{macro} + R_{macro}} \quad (12)$$

Macro F1 score, first calculate the F1 score of each class when it is a positive class, and then average all the F1 scores. First, we average the sum of the categories:

$$P_{macro} = \frac{\sum_{i=1}^{n} P_i}{n} \quad R_{macro} = \frac{\sum_{i=1}^{n} R_i}{n} \quad (13)$$

Then, the F1 value calculated by using the F1 calculation formula is Macro-F1[5].

$$F1_{macro} = 2 \cdot \frac{P_{macro} \cdot R_{macro}}{P_{macro} + R_{macro}} \quad (14)$$

By adjusting parameters and calculating the accuracy and other generalization indicators of the model according to formulas (x) to (X), the results are shown in Table 2.

**Table 2** Generalization ability analysis and accuracy test of Bayesian-based logistic regression multi-classification model

| Evaluation index | 0 is the positive | 0.3266 is positive | 0.6734 is positive | 1 is positive | MacroAVG | MicroAVG |
|---|---|---|---|---|---|---|
| true_positive | 869 | 2399 | 2426 | 1000 | 1679.5 | 1679.5 |
| false_positive | 23 | 270 | 238 | 25 | 36.5 | 136.5 |
| false_negative | 238 | 26 | 24 | 259 | 136.5 | 136.5 |
| true_negative | 6089 | 4602 | 4610 | 6003 | 5298.5 | 5297.5 |
| precision | 0.969 | 0.911 | 0.911 | 0.996 | 0.942 | 0.916 |
| sensitivity | 0.791 | 0.988 | 0.989 | 0.802 | 0.892 | 0.916 |
| specificity | 0.987 | 0.951 | 0.949 | 0.977 | 0.981 | 0.975 |
| accuracy | 0.913 | 0.932 | 0.931 | 0.919 | 0.919 | 0.931 |
| F-measure | 0.856 | 0.938 | 0.955 | 0.869 | 0.912 | 0.931 |

At the same time, the ROC curves and different AUC values of 4 different positive class

cases are calculated, as shown in Figure 2:

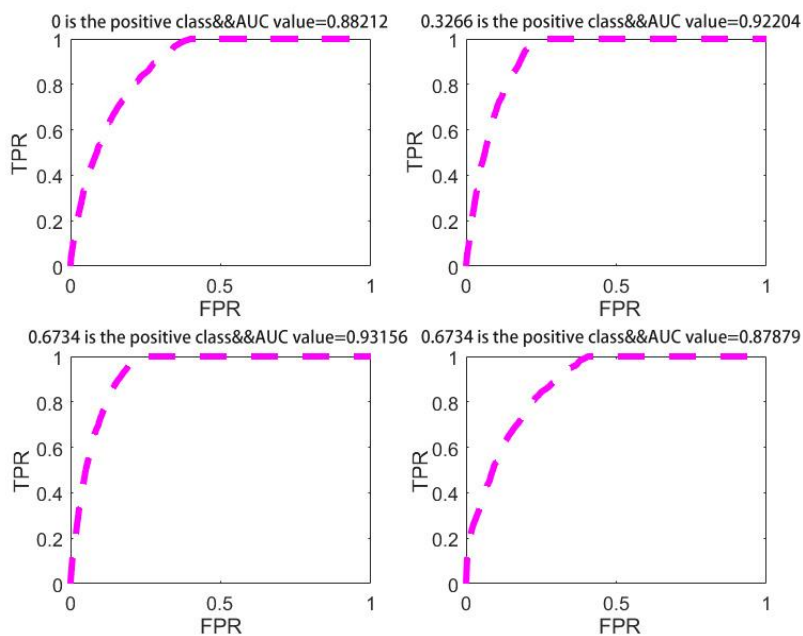

Figure 2 is based on the Bayesian multi-classification model ROC curve

## 4.4 Results of the multiple time scale model calculation

Taking 1701 matches as an example, under all rounds, through the model constructed above, 1701 matches (the first 10 samples) and the probability of belonging to a certain category are predicted[6], as shown in Table 3 below:

**Table 3** The probability that a sample belongs to a certain class

| The probability of prediction is 0 | The probability of prediction is 0.3266 | The probability of prediction is 0.6734 | The probability of prediction is 1 | Final forecast result |
|---|---|---|---|---|
| 7.19E-15 | 0.859431 | 9.17E-23 | 0.140568 | 0.3366(Player 2 wins) |
| 1 | 7.01E-17 | 9.05E-17 | 7.39E-41 | 0(Player 2 wins) |
| 1.55E-08 | 3.25E-28 | 0.99997 | 4.95E-05 | 0.6734(Player 1 wins) |
| 2.24E-13 | 0.671197 | 1.28E-19 | 0.328808 | 0.3266(Player 2 wins) |
| 7.55E-15 | 0.875219 | 7.88E-23 | 0.124783 | 0.3266(Player 2 wins) |
| 3.31E-14 | 0.999998 | 7.88E-39 | 3.68E-06 | 0.3266(Player 2 wins) |
| 1.60E-16 | 4.39E-53 | 1 | 2.69E-18 | 0.6734(Player 1 wins) |
| 3.99E-14 | 0.999998 | 1.50E-38 | 4.41E-06 | 0.3266(Player 2 wins) |
| 1.09E-22 | 5.32E-22 | 4.72E-22 | 1 | 1(Player 1 wins) |
| 0.222249 | 1.32E-17 | 0.777851 | 8.36E-17 | 0.6734(Player 1 wins) |

# 5 Problem 2: Momentum evaluation model based on trend testing

## 5.1 "Momentum" definition

We defined momentum by comprehensively considering the fluctuation trend of momentum. We took the scoring situation of the three rounds before and after the game and the seven rounds before and after respectively, and gave different weights to these two parts. In order to more accurately capture the momentum changes in the game, the impact of continuous scoring is considered. The mathematical model is as follows: Let $M(n)$ represent the scoring situation of the three rounds before and after the nth round, that is, in the round before and after the nth round the total score. Let $N(n)$ represent the scoring situation of the seven games before and after, that is, the total score in the first 3 rounds and the last 3 rounds of the nth round. Since the number of recent rounds has a greater impact on momentum, we introduce weights $a_1$ and $a_2$ to meet the following conditions: $a_1 + a_2 = 1$

Therefore, the momentum score $P(s)$ at the $s$-th round is calculated as follows:

$$P(n) = a_1 * M(n) + a_2 * N(n) \tag{15}$$

In the formula, $w_1$ and $w_2$ are different weights given to balance the scoring situation of the three games before and after and the seven games before and after. When considering that player 1 continuously scores, the score calculation for the three games before and after and the seven games before and after is as follows:

$$\begin{cases} M_1(n) = \dfrac{\sum_{s-1}^{s+1} r + \alpha_1 \cdot e^{2k}}{3} + 0.5 \\ N_1(n) = \dfrac{\sum_{s-3}^{s+3} r + \beta_1 \cdot e^{k}}{7} + 0.5 \end{cases} \tag{16}$$

In the formula, $\sum_{s-1}^{s+1} r$ represents the total score in the round before and after round $s$, and $\sum_{s-3}^{s+3} r$ represents the total score in the three rounds before and after round $s$. This situation Regardless of the continuous scoring situation, $r$ is the corresponding score, and only the total number of scores is counted. As the ordinary score part, it is calculated based on the number of non-consecutive scoring rounds. When counting three games, it is considered that the importance of consecutive scoring is greater than seven games, so $e^{2k}$ serves as the growth scalar for three games, $e^k$ serves as the growth scalar for seven games,

$$e^{2k} > e^k$$

$\alpha_1 \cdot e^{2k}$ and $\beta_1 \cdot e^k$ represent continuous scoring. As an additional part, the momentum score is increased according to the consecutive scoring rounds, affecting the overall "momentum" change. $\alpha_1$ and $\beta_1$ represent the change (growth) factor of continuous scoring in different statistical round intervals. Different growth factor values are assigned within, $k$ is the number of consecutive scoring games, $0 \leq k \leq 7$.

Since player 1 and player 2 are independent individuals, when calculating the total momentum score, the value of the momentum part is relative. Therefore, when player 1 scores consecutively and player 2 loses consecutive points, player 2 has three games before and after. The score calculation for the seven games is as follows:

$$\begin{cases} M_2(n) = \dfrac{\sum_{s-1}^{s+1} r + \alpha_2 \cdot e^{2k}}{3} + 0.5 \\ N_2(n) = \dfrac{\sum_{s-3}^{s+3} r + \beta_2 \cdot e^{k}}{7} + 0.5 \end{cases} \tag{17}$$

When player 2 loses consecutive points, $\alpha_2 \cdot e^{2k}$ and $\beta_2 \cdot e^{k}$ represent the continuous ten-point part. As an additional part, the momentum score is reduced according to the number of consecutive rounds of losing points, affecting the change of the overall "momentum". $\alpha_2$ and $\beta_2$ represent the change of consecutive points. (Attenuation) factor, assigning different attenuation factor values in different statistical round intervals, $k$ is the number of consecutive scoring games, $0 \leq k \leq 7$.

It can be seen from this that player 1 and player 2 are opposed to each other, satisfying

$$\begin{cases} \alpha_1 + \alpha_2 = 0 \\ \beta_1 + \beta_2 = 0 \end{cases} \tag{18}$$

$$0 \leq P(n) \leq 1 \tag{19}$$

## 5.2 Use the "momentum" score as the output and substitute it into the model in question 1 for training.

In summary, by reasonably setting the weights and change factors to better reflect changes in momentum, a more accurate momentum score can be calculated.

In this article, we assign $\alpha_1 = 0.0012$, $\beta_1 = 0.0025$, $w_1 = 0.7$, $w_2 = 0.3$. Based on the calculation method of "momentum" score mentioned above, we calculated the momentum score of each round in all existing rounds, and used it as our problem to build a multi-time scale winning rate classification prediction model based on Bayesian probability. Carry out model training and improve the accuracy by adjusting parameters and changing the structure of the network. Finally, the comparison of the "momentum" score changes in the training set and the test set is shown in Figure 3.

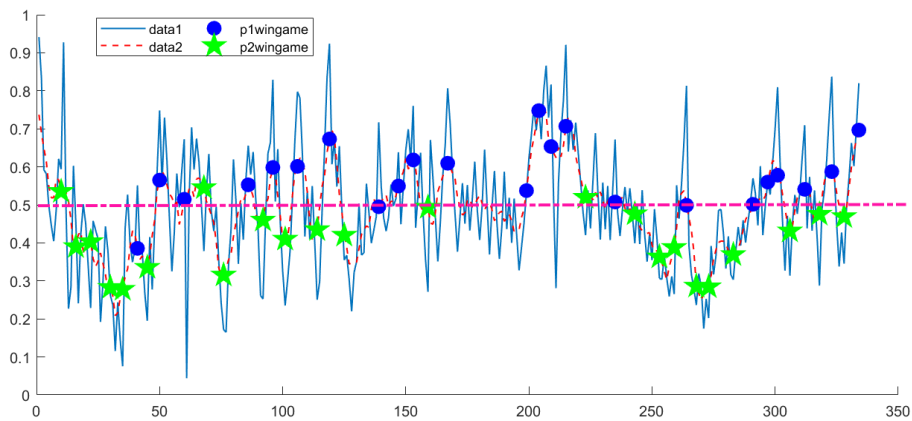

**Figure 3** Training results of the model after changing the "momentum" score as the output

## 5.3 Unsupervised "momentum" evaluation model of AHP analytic hierarchy process based on subjective evaluation.

The quantitative scale of the nine-scale method is as shown in the table. The nine-scale method performs a pairwise comparison of the importance of elements at the same level in the hierarchical structure diagram, thereby constructing the required judgment matrix:

Table 4 Quantity scaling using the nine-scale method

| scale | definition |
|---|---|
| 4 | $a_i$ is equally important as $a_j$ |
| 5 | $a_i$ is slightly more important than $a_j$ |
| 6 | $a_i$ is obviously more important than $a_j$ |
| 7 | $a_i$ is much more important than $a_j$ |
| 8 | $a_i$ is extremely important than $a_j$ |
| 3-1 | $a_i$ is more important than $a_j$ (inverse comparison) |

**Step 1:** Establish a criterion-level judgment matrix

$$R = \begin{bmatrix} r_{11} & r_{12} & \cdots & r_{1n} \\ r_{21} & r_{22} & \cdots & r_{2n} \\ \cdots & \cdots & \ddots & \vdots \\ r_{n1} & r_{n2} & \cdots & r_m \end{bmatrix} = \begin{bmatrix} 1 & r_{12} & \cdots & r_{1n} \\ \dfrac{1}{r_{12}} & 1 & \cdots & r_{2n} \\ \cdots & \cdots & \ddots & \vdots \\ \dfrac{1}{r_{1n}} & \dfrac{1}{r_{2n}} & \cdots & 1 \end{bmatrix} \quad (20)$$

**Step 2:** Weight of criterion layer to target layer

$$W_i = \frac{\sum_{j=1}^{n} a_{ij} + \frac{n}{2} - 1}{n(n-1)} \quad (21)$$

**Step 3:** Weight value

$$\begin{cases} I(A, W^*) = \dfrac{1}{n^2} \sum_{i=1}^{n} \sum_{j=1}^{n} |\alpha_{ij} + w_{ij} - 1| \\ W^* = (w_{ij})_{n*n} \\ w_{ij} = \dfrac{w_i}{w_i + w_j} \end{cases} \quad (22)$$

**Step 4:** Establish a solution-level judgment matrix

**Step 5:** Check the consistency of the weight vector and the combined weight vector

Assume that the pairwise comparison judgment matrix of related factors in layer B is tested for consistency in a single sorting, and the single sorting consistency index is obtained as CI(j), (j=1,...,m), and the corresponding average random The consistency index is RI(j), (CI(j), RI(j) have been obtained during single-level sorting), then the total sorting random consistency ratio of layer B is:

$$CR = \frac{\sum_{j=1}^{m} CI(j) a_j}{\sum_{j=1}^{m} RI(j) a_j} \quad (23)$$

When the CR value is less than 0.1, we believe that the total hierarchical ranking results have satisfactory consistency and accept the analysis results. Input the preprocessed data into the above model, and the single ranking CR of the result calculated by Matlab is 0.085, which is less than 0.1, indicating that the results are consistent.

Use Matlab to perform programming and sort the cities corresponding to different pipelines. The final results are shown in Table 5.2. Due to limited space, we only show the scores and rankings of the top 10.

**Table 5** Topsis "Momentum" evaluation scores and rankings (first 10 rounds)

| Number of rounds | score | standardization | Ranking |
| --- | --- | --- | --- |
| Round 1 | 0.4329 | 0.0025 | 202 |
| Round 2 | 0.4931 | 0.0028 | 204 |
| Round 3 | 0.5047 | 0.0029 | 201 |
| Round 4 | 0.4928 | 0.0028 | 203 |
| Round 5 | 0.5045 | 0.0028 | 200 |
| Round 6 | 0.4958 | 0.0028 | 230 |
| Round 7 | 0.4927 | 0.0028 | 227 |
| Round 8 | 0.4968 | 0.0028 | 226 |
| Round 9 | 0.4845 | 0.0027 | 235 |
| Round 10 | 0.4856 | 0.0027 | 228 |

The corresponding "momentum" score visualization is shown in Figure 4:

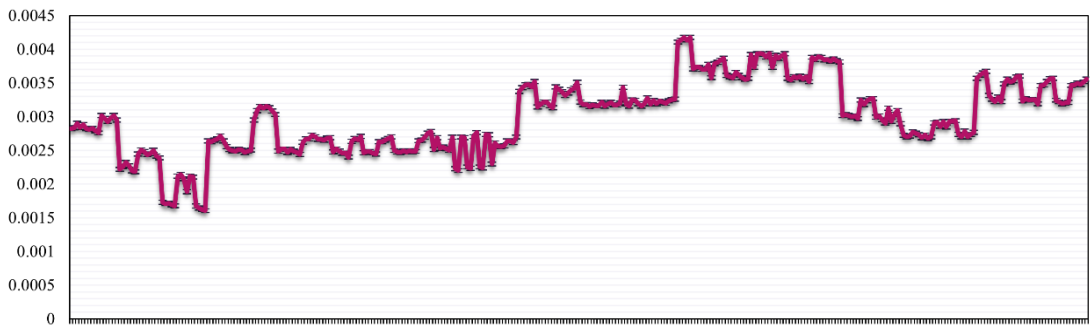

**Figure 4** "Momentum" score changes during the game

## 5.4 Trend correlation test based on cosine similarity

We first analyze the three-dimensional surface fitting of the seven indicators, "momentum" score and winning rate, and fit it through the ploy22 function[4]. The function model is: $p00+p10*x+p01*y+p20*x^2+p11*x*y+p02*y^2$. At the same time, the corresponding surface fitting formula and R-squared (R2) value (part) are obtained as follows:

Linear model Poly22:

    $model(x,y) = p00 + p10*x + p01*y + p20*x^2 + p11*x*y + p02*y^2$

    Coefficients(with 95% confidence bounds):

        p00 =   0.07587   (-0.03598 , 0.1877)
        p10 =   0.4065    (0.05124 , 0.7618)
        p01 =  -0.04929   (-0.4591 , 0.3605)
        p20 =   0.147     (-0.1491 , 0.443)
        p11 =   0.6151    (0.2775 , 0.9527)

p02 =   0.3154    (-0.1003 , 0.7311)

The R-squared(R2) value is 0.7599

Euclidean Distance: Also known as straight-line distance, it is the most common distance measurement method. The Euclidean distance between two points and $(x_2, y_2)$ is calculated as follows:

$$\textbf{Euclidean Distance} = \sqrt{(x_2 - x_1)^2 + (y_2 - y_1)^2} \tag{24}$$

This formula actually calculates the straight-line distance between two points. The smaller the Euclidean distance is, the more similar the two vectors are. The final result of cosine similarity calculated through Matlab is: 0.0923.

# 6 Problem 3: Momentum trend analysis model

## 6.1 Problem Analysis

Starting from two directions, the characteristics of the game and the volatility of momentum, a mathematical model is built to predict when the momentum may change in the game.

First, from the perspective of game characteristic analysis, a series of quantifiable variables are defined, and factors related to momentum changes are identified through statistical analysis. Momentum transition points can be predicted by setting thresholds on historical data, and then a classification algorithm is used to build a model to predict the probability of momentum transitions based on current game characteristics. On the other hand, from the perspective of momentum volatility, we focus on quantitative indicators of game momentum and analyze the fluctuation trends of these indicators through time series. This includes observing momentum change patterns after key scores, identifying fluctuation patterns through time series forecast models or sequence analysis methods, and predicting future momentum changes.

Combining these two analytical methods, we build a comprehensive model that combines match characteristics with momentum volatility to provide a comprehensive perspective to predict possible changes in match flow. Through training and validation on historical data and testing on real-time data, we evaluate model accuracy and make adjustments based on actual performance. This model not only provides coaches and players with a basis for tactical adjustments, but also helps make smarter decisions at critical moments and better grasp the momentum of the game.

## 6.2 "Momentum" scores and changes in wins are mostly trend analysis models

By analyzing the changing trend of winning probability over time for the full layout and all players 1 under 1701 games and the evolution of the game schedule, as shown in Figure 5:

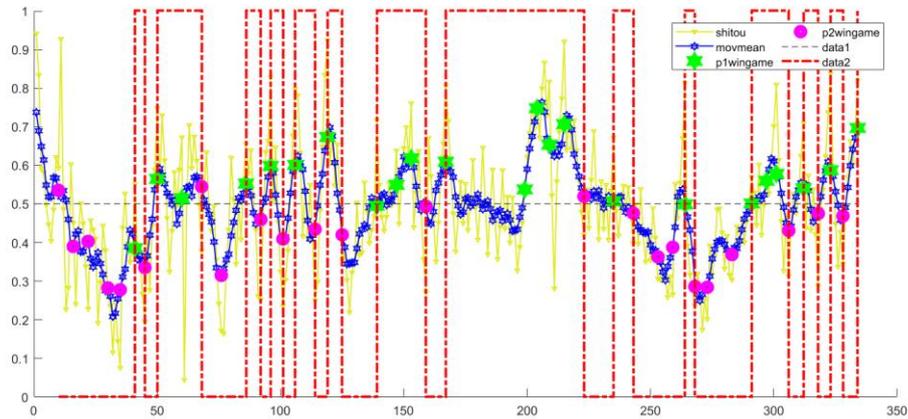

**Figure 5** 1701 Analysis of player 1's "momentum" score and winning trend in games

can be seen from the picture that at the beginning of the game, Djokovic remained strong and continued to suppress Alcaraz, causing the latter to gradually lose his advantage. The green dot indicates the victory of Djokovic, who defeated Alcaraz 6:1 in the first set.

Analysis shows that when the two sides in the game are similar in strength, the momentum will fluctuate and reverse, making it difficult for one side to completely suppress the other. Although Alcaraz came back quickly in the second set, the match remained tight. This fierce duel showed an obvious feature, that is, the victory points of all rounds were close to the peak of momentum, and there was an obvious trough after the peak. After Alcaraz won the second set, the average volatility momentum increased to above 0.5, showing the impact of key games on the overall momentum average.

The game continued to unfold, and after the breakout, Alcaraz realized a period of high momentum. However, the situation was reversed in the next set, highlighting the impact each set restart has on the momentum of both players, with performance in the previous set or previous sets being one of the factors.

An analysis of the situation at the beginning and end of the game found that in the fourth set, Alcaraz was once again at a disadvantage. However, in the fifth set, the two players once again fell into a fierce battle. The fluctuations in momentum are similar to those in the second set, indicating that the volatility of momentum itself can unearth some characteristics and become useful indicators to guide players to win in the next set or the entire game.

Therefore, there are two ideas for analyzing the problem. First, start with the known game characteristics and study the response of player momentum when the game characteristics change. Coaches can capture key indicators and predict game trends. Secondly, start from the volatility of momentum. Momentum fluctuations have specific responses to the previous fluctuations. If one party takes the advantage after a stalemate, it is likely to have a larger momentum state and achieve multi-game suppression. This kind of analysis helps coaches or players better judge the current situation and formulate reasonable strategies.

## 6.3 Momentum fluctuation prediction model based on network search method under the condition that players serve successively

On this basis, we considered the comparison of score difference, psychological factors, personal technical ability, running distance, set difference and momentum score when players served first and second. We incorporate these factors into the model for question 2 for training. By perturbing different momentum indicators, we observe the performance of momentum in continuous changes, that is, the outcome of the game, and analyze the turning points of score changes to provide targeted suggestions. The player. When only considering the change of a certain indicator, we use the network search method to conduct sensitivity analysis and observe the impact of the indicator on the momentum under the same step change.

Taking the score difference as an example, it can be seen from the figure that as the score difference decreases, both curves show an upward trend, indicating that the momentum score increases as the score difference decreases. When the game point difference decreases, the momentum score increases, indicating that the game is more tense and uncertain, and whether the team serving first or second will respond actively.

For player 1 to serve first, the momentum score is lower when the score difference is large, indicating that the side with the greater advantage has less momentum. As the score difference decreases and approaches 0, the momentum score increases, indicating that the disadvantaged side is gaining momentum. For players who serve after 1, the momentum score also increases when the score difference decreases, but the overall increase is smaller.

In terms of the influence of psychological factors, when player 1 serves first, the psychological factors rise rapidly within a certain range, and then decrease as the psychological factors increase or decrease after reaching the peak. This suggests that there is a psychological "sweet spot" in which Player 1 has the strongest momentum to serve first. In contrast, the psychological factors of player 1's post-serve are less sensitive to the influence of momentum, showing a gentle upward and downward trend.

Focus on analysis from three perspectives:

1. Intersection point: From the perspective of player 1, the blue line indicates the second serve, and the pink line indicates the first serve. Overall, as the psychological factors change, the momentum of successive serves is quite different, but when the psychological factor is about 0.06, the momentum of the two is equal. This shows that under this psychological factor value, the momentum remains unchanged when player 1 serves successively.

2. Average momentum: The green line indicates the change in the overall momentum chart after averaging the momentum of successive serves. It can be known from the trend of average momentum that player 1's average momentum reaches the maximum value near the intersection point of the momentum of serving successively, and when it is greater than 0.5, the momentum is the largest regardless of whether the serve is served successively under this psychological factor state.

3. Momentum trend: observe the changes in the absolute value of momentum when the momentum is greater than 0.5 and less than 0.5. Since the tennis match is a continuous serve in a certain game, it will be replaced by the opponent's continuous serve when moving to the next level. Therefore, with the enhancement of psychological factors, the winning rate of the game corresponding to the late serve is higher. After the intersection point, as psychological factors improve, the probability of winning by serving later gradually weakens, while the probability of winning by serving first increases.

In general, by observing the momentum change of other indicators such as psychological factors, it can effectively provide tactical suggestions and guidance for tennis matches. Next, we will simultaneously consider the impact of changes in the two indicators on the momentum score, which is shown in Figure 6 through visual analysis.

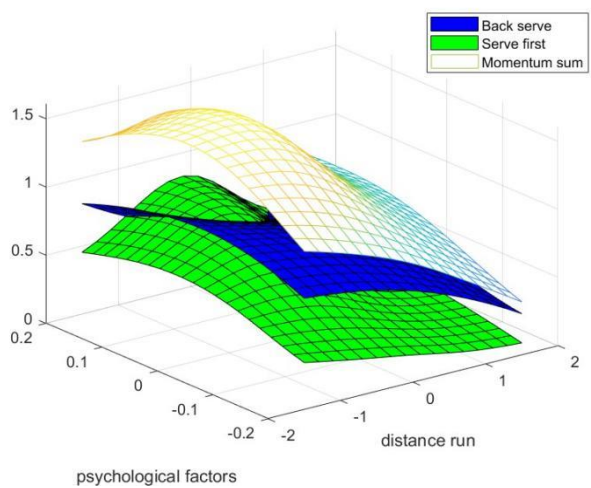

**Figure 6** three-dimensional comparative changes of two different indicators and momentum scores

Using this 3D plot, we see that momentum changes as a function of running distance and psychological factors. When these two characteristics make the momentum of the first serve and the second serve basically equal, the momentum can maintain a high value in both the first serve and the second serve situations. For the players, their advantage is still obvious in every round and after changing serves. Judging from the image characteristics, the maximum momentum mean is distributed near the intersection of the two surfaces of the first serve and the second serve, which is basically consistent with the one-dimensional situation. This shows that to determine whether a player is in an advantageous situation, we can determine the player's state change by whether the model results are dominated by the advantage of serving first.

## 6.4 Trend analysis based on wavelet transform

Wavelet transform is used to analyze the time changes of the momentum score signal at different frequencies. In wavelet analysis, the momentum score signal is converted into time-frequency space so that the time and frequency information of the signal can be observed simultaneously. The basic mathematical expression of wavelet transform is:

$$W(a,b) = \int_{-\infty}^{\infty} f(t) \cdot \frac{1}{\sqrt{|a|}} \cdot \varphi \cdot \left(\frac{t-b}{a}\right) dt \qquad (25)$$

The amplitude time-frequency diagram of the wavelet transform is obtained, as shown in Figure 7.

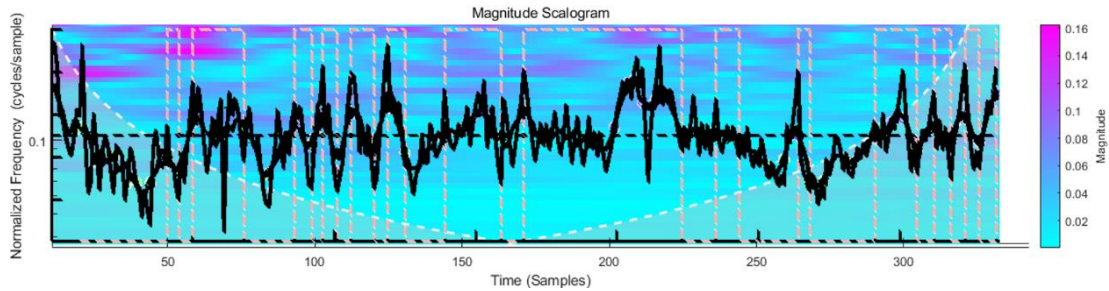

**Figure 7** Trend analysis of wavelet transform

The horizontal axis represents the change in game rounds over time, and the vertical axis represents the normalized momentum score. The dark areas in the figure represent the amplitude of the wavelet coefficients, that is, the energy or importance of the signal at a specific time and frequency. The illustrated amplitude time-frequency plot shows the strength of the momentum score signal at different frequencies and time points. Color bars represent the amplitude of the wavelet transform coefficients, and dark colors represent strong signals at the corresponding scale and time point. The image highlights moments or phases of the game when Player 1 had strong momentum[7].

# 7 Establishment and solution of model for question 4

## 7.1 problem analysis

Question four requires validating the model built in the first three questions, testing it through one or more games, and evaluating the model's generalization ability and transferability. We collected relevant data and other competition datasets and fed them into the previously constructed classification prediction model. Evaluate the model's migration performance by observing how the predictions match actual momentum values. The analysis and modeling process of Question 4 is shown in the figure 8 below.

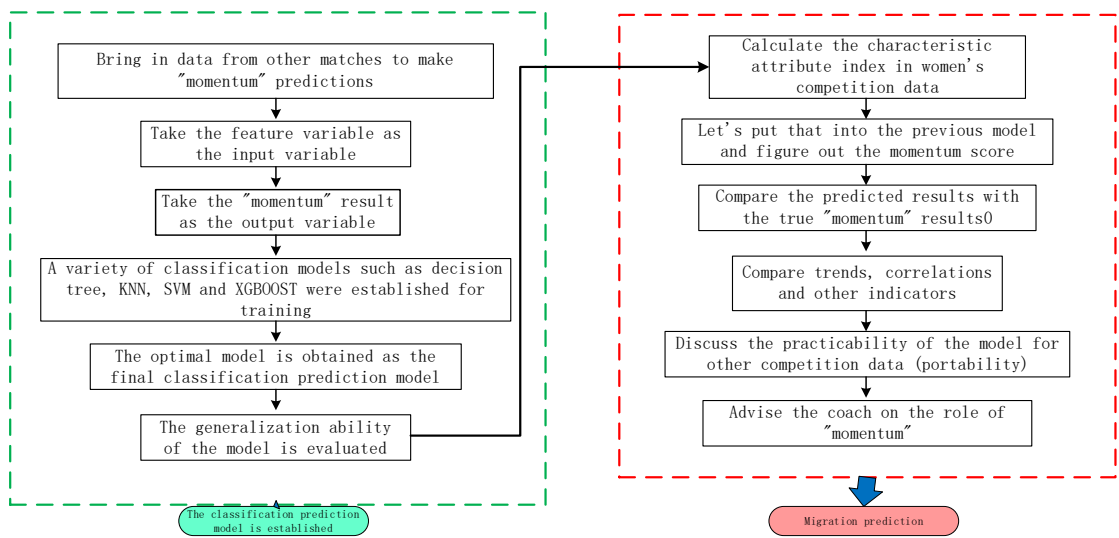

**Figure 8** Analysis and modeling process for question 4

## 7.2 Model prediction and generalization ability verification

Using other models in the existing data set to evaluate the generalization ability, the training results of the model and the fitting situation of the momentum changes are shown in Figure 9.

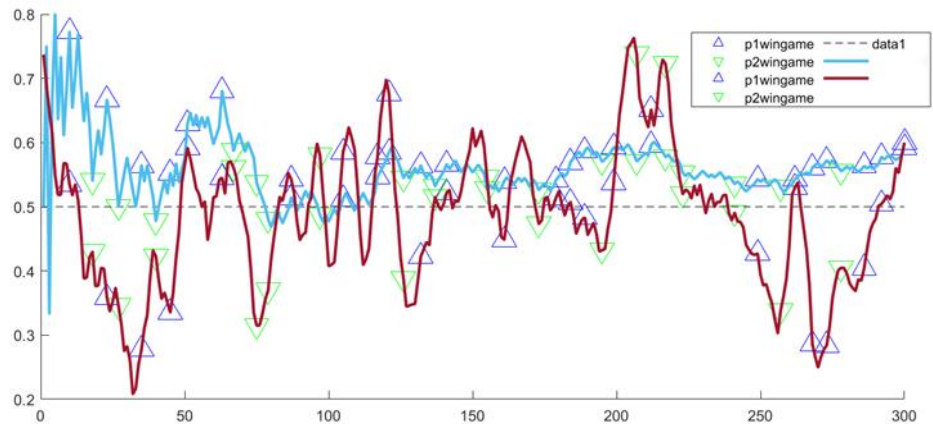

**Figure 9** Comparison of momentum model prediction results for other games

It can be observed from the figure that, firstly, for the momentum results we found in other data sets, comparing with the momentum results in our data set, there is a significant similarity in the trend, but a clear difference in the magnitude of the fluctuations. Despite this difference, both data sets exhibit the ability for out wins to be correlated with momentum spikes or troughs.

This suggests that a sudden increase or maximum in momentum often indicates a possible win. This annotation is very accurate for the dataset we have trained on. However, for new competitions, the annotation accuracy has decreased, and some inevitable incorrect annotations may occur.

The reason for this error is that we used the features used in the game when training the data, that is, we ignored the personal factors and relative ability differences between the two players. To train the model more accurately, we need more data about players' personal abilities, such as winning rate, second serve rate, net hitting rate and other comprehensive indicators, in order to more accurately extract the momentum change characteristics of players. Or we can learn enough features about that player's games against any one opponent to generate a model about a single player that, taken together, can predict the outcome of a game more accurately.

## 7.3 Use the data of female tennis players to predict the migration of the model.

In the newly collected female tennis data, we predict the migration of the model. We know the real momentum value. First, we train the model in the new data set and output The training effect and corresponding prediction results of the new momentum numerical model are shown in Figures 10 and 11 respectively.

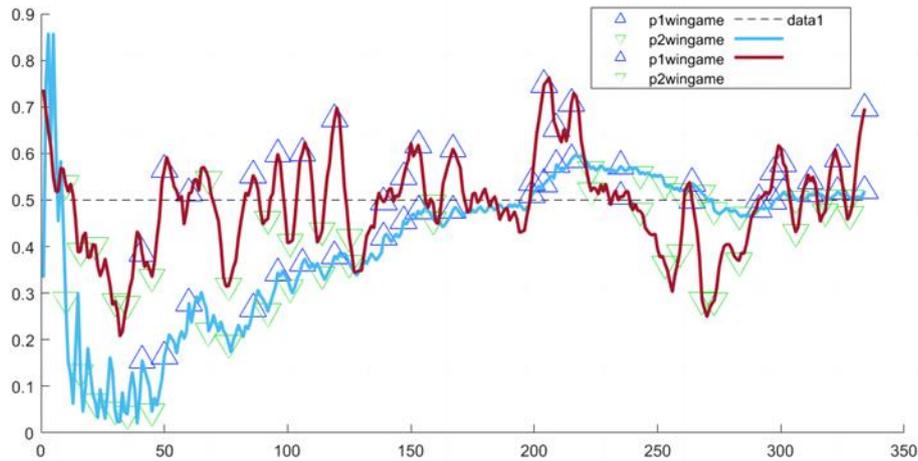

**Figure 10** Fitting of the training set and test set of the model

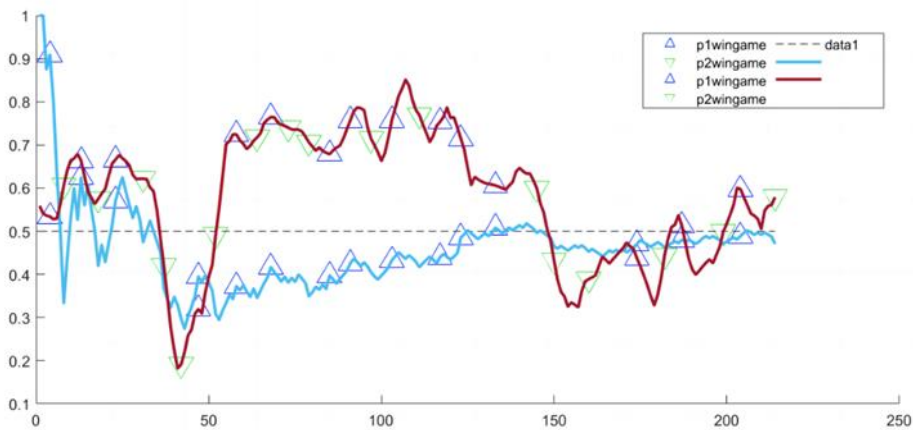

**Figure 11** Bringing in girl data to model migration prediction momentum fitting effect

According to the illustration, if we apply it to different types of competitions, such as women's tennis competitions, since the competition system is different from men's, using a best-of-three-set system, this may be related to the physical performance of female athletes. In training this model, we use the men's model but normalize it to data specific to the women's game. The final training results can well reflect the relationship between momentum extremes and game victory, which shows that our model has good generalization. By adjusting its internal parameters, the model is able to accurately predict a player's likely winning situations or changes in momentum during a match.

## 8 letter of advice for question 5

Our research indicates that momentum is a crucial factor influencing match outcomes, manifested in scoring, technical execution, and the pace of the game. Coaches should focus on training to simulate players' psychological resilience and concentration when trailing in scores or during crucial serving moments. Data analysis reveals a significant impact of the serving phase on match momentum, suggesting that coaches should concentrate on serving and receiving training to enhance stability and aggression.

Additionally, coaching teams can utilize statistical data analysis of match strategies and player performance patterns to anticipate opponents' performances under different score pressures. In terms of game strategy, coaches should guide players to adjust strategies flexibly in response to momentum changes while maintaining their playing style, such as reducing errors and adopting a conservative strategy when observing a decline in momentum, or actively seeking attacking opportunities when momentum is on the rise.

Lastly, emphasizing players' adaptability in matches is crucial, cultivating their ability to instantly recognize and respond to various situations, rapidly adjusting their mindset and strategy. In conclusion, momentum plays a key role in tennis matches and serves as a core consideration in tactical preparation and match execution. Through scientific training and data analysis, players can better grasp the rhythm and dynamics of the game, gaining an advantage.

In sports competitions, momentum is a constantly changing force that impacts the outcome. Players should be thoroughly prepared, establishing psychological and strategic readiness through opponent analysis and training. Psychological and physical preparation is equally vital, requiring the cultivation of mental resilience, composure, and focus. Game strategies should be flexible, adjusted according to changes in momentum, assessing the situation, and changing tactics when necessary. Players need to demonstrate unwavering perseverance in adversity, persisting even when faced with unfavorable momentum.

A player's success depends not only on technical and tactical skills but also on how they manage intangible factors in the game, such as momentum and pressure, and utilize team support to optimize performance.

Overall, the wavelet transform's time-frequency graph can analyze the patterns of players' success rates throughout the course of the game. However, correlating this information with actual game data requires more specific knowledge of the game and contextual information. Wavelet analysis is a powerful tool for capturing the non-stationary characteristics of signals, particularly useful in analyzing dynamics in sports competitions.

## Conflict of Interest

Authors have no conflict of interest to declare.